# Using JAGS for Bayesian Cognitive Diagnosis Modeling: A Tutorial

Peida Zhan (Zhejiang Normal University)[1]

Hong Jiao (University of Maryland, College Park)

Kaiwen Man (University of Maryland, College Park)

**Note:**

This is an arXiv preprint, may not be the final version. For reference:

[1] Corresponding author: Peida Zhan, Department of Psychology, College of Teacher Education, Zhejiang Normal University, No. 688 Yingbin Road, Jinhua, Zhejiang, 321004, P. R. China. Email: pdzhan@gmail.com

# Using JAGS for Bayesian Cognitive Diagnosis Modeling: A Tutorial

## Abstract

In this article, the **JAGS** software program is systematically introduced to fit common Bayesian cognitive diagnosis models (CDMs), including the deterministic inputs, noisy "and" gate (DINA) model, the deterministic inputs, noisy "or" gate (DINO) model, the linear logistic model, the reduced reparameterized unified model (rRUM), and the log-linear CDM (LCDM). The unstructured latent structural model and the higher-order latent structural model are both introduced. We also show how to extend those models to consider the polytomous attributes, the testlet effect, and the longitudinal diagnosis. Finally, an empirical example is presented as a tutorial to illustrate how to use the **JAGS** codes in **R**.

## Introduction

In recent years, many cognitive diagnosis models (CDMs) have been proposed, such as the deterministic inputs, noisy "and" gate (DINA) model (Haertel, 1989; Junker & Sijtsma, 2001; Macready & Dayton, 1977), the deterministic input, noisy "or" gate (DINO) model (Templin & Henson, 2006), the linear logistic model (Maris, 1999), and the reduced reparameterized unified model (rRUM; Hartz, 2002). Some general CDMs are also available, such as the log-linear CDM (LCDM; Henson, Templin, & Willse, 2009), the generalized DINA model (de la Torre, 2011), and the general diagnosis model (von Davier, 2008).

As the advance of computing power and the Markov chain Monte Carlo (MCMC) algorithms, Bayesian cognitive diagnosis modeling has become increasingly popular (e.g., Culpepper, 2015a; Culpepper & Hudson, 2017; DeCarlo, 2012; de la Torre & Douglas, 2004; Huang & Wang, 2013; Li, Cohen, Bottge, & Templin, 2016; Li & Wang, 2015; Sinharay & Almond, 2007; Zhan, Jiao, & Liao, 2018; Zhan, Jiao, Liao, & Bian, in press). Multiple software programs are available to implement some Bayesian MCMC algorithms, such as **WinBUGS** (Lunn, Thomas, Best, & Spiegelhalter, 2000), **OpenBUGS** (Spiegelhalter, Thomas, Best, & Lunn 2014), **JAGS** (Plummer, 2017), and **Mcmcpack** (Martin, Quinn, & Park, 2011) package in **R** (R Core Team, 2016). However, to date, there is still a lack of systematic introduction to using such software programs to fit Bayesian CDMs.

Unlike the frequentist approach which treats model parameters as fixed, the Bayesian approach considers them as random and uses (prior) distributions to model our beliefs about them. Within the frequentist framework, parameter estimation refers to a point estimate of each model parameter. By contrast, in a fully Bayesian analysis, we seek a whole (posterior) distribution of the model parameter, which includes the entire terrain of peaks, valleys, and plateaus (Levy & Mislevy, 2016). The posterior distribution of parameters given the data is proportional to the product of the likelihood of the data given the parameters and the prior distribution of the parameters. Typically, the posterior distribution is represented regarding the posterior mean (or median, or mode) as a summary of central tendency, and the posterior standard deviation as a summary of variability.

The advantages of adopting a Bayesian MCMC estimation over a frequentist estimation, e.g., maximum likelihood estimation (see, Wagenmakers, Lee, Lodewyckx, & Iverson, 2008) include (a) it does not depend on asymptotic theory; (b) it treats both item and person parameters as random effects; (c) it incorporates the principle of parsimony by marginalization of the likelihood function, and (d) it is more robust in handling complex models. Also, in Bayesian estimation, the percentiles of the posterior can be used to construct the credible interval (or Bayesian confidence interval) which can be used for testing of significance (Box & Tiao, 1973). Further, it is also easy to conduct model-data fit with posterior predictive model checking (PPMC).

Currently, many CDM studies are quite technical and limited to statistical and psychometric researchers (Templin & Hoffman 2013). There is a lack of available software for more applied practitioners who would like to use CDMs in developing their diagnostic testing programs or conducting empirical research. Moreover, most existing programs for cognitive diagnosis are either limited with model options or commercialized. For example, the **Arpeggio Suite** (Bolt et al., 2008), the **mdltm** (von Davier, 2005), the **CDM** package (George, Robitzsch, Kiefer, Gross, & Uenlue, 2016), and the **GDINA** package (Ma & de la Torre, 2016b) limit users to a few options. In addition, albeit the **Mplus** (Muthén & Muthén, 2010) and the **flexMIRT** (Cai, 2017) can be used to fit many CDMs (e.g., Hansen, Cai, Monroe, & Li, 2016; Templin & Hoffman, 2013), their commercialization prevents unauthorized users especially students from accessing these software without purchasing.

In this article, we demonstrate how to use the freeware, **JAGS**, to fit several popular CDMs and present the code. It is expected that the researchers can adapt the code to fit extended CDMs,which cannot be fitted in existing software or packages for their   research or application purposes.

In general, **JAGS** makes it easy to construct a Markov chain for parameters. It does not require users to derive the posterior distribution of the model parameters by hand. Movereover,, the **R2jags** package (Version 0.5-7; Su & Yajima, 2015) in **R** could be easily used to call the **JAGS**. Furthermore, It should be noted that the **JAGS** code presented in this study can be generalized easily to other BUGS software programs by minor editing including **WinBUGS** and **OpenBUGS**[2].

The following sections first illustrate **JAGS** codes for five CDMs: (1) the DINA model; (2) the DINO model; (3) the LLM; (4) the rRUM, and (5) the LCDM. Besides those five models, which are based on the unstructured (or saturated) latent structuralels, the higher-order latent structural model (de la Torre & Douglas, 2004) is also demonstrated. Further, the extensions to the polytomous attributes, the testlet effect, and the longitudinal diagnosis using **JAGS** are presentedas well. Lastly, an empirical example analysis is conductedto illustrate how to use the **R2jags** package to run the **JAGS** code.

## The DINA Model

Let $Y_{ni}$ be the response of person $n$ ($n = 1,..., N$) to item $i$ ($i = 1,..., I$). Let $\alpha_{nk}$ be the binary variable for person $n$ on attribute $k$ ($k = 1,..., K,$), where $\alpha_{nk} = 1$ indicates that person $n$ shows mastery of attribute $k$ and $\alpha_{nk} = 0$ indicates non-mastery, and let $\boldsymbol{\alpha}_n = (\alpha_{n1}, \ldots, \alpha_{nK})'$ be the $n$-th person's attribute pattern. Let $q_{ik}$ denote the element in a $I$-by-$K$ Q-matrix (Tatsuoka, 1983), with $q_{ik}$ indicating whether attribute $k$ is required to answer item $i$ correctly. If the attribute is required, $q_{ik} = 1$, otherwise, $q_{ik} = 0$.

---

[2]  Some tiny differences between JAGS and OpenBUGS (or WinBUGS) can be found in the manual of JAGS (Plummer, 2017).

Among CDMs, the DINA model is one of the most popular because of its simple structure and straightfoward interpretation. The DINA model can be expressed as:

$$p_{ni} = P(Y_{ni} = 1 \mid \boldsymbol{\alpha}_n) = g_i + (1 - s_i - g_i)\eta_{ni}, \qquad (1)$$

where $p_{ni}$ is the correct response probability of person $n$ to item $i$; $s_i$ and $g_i$ are the slipping and guessing probability, respectively, of item $i$, which describe the item-level aberrant response probability; $\eta_{ni}$ is called the ideal response for person $n$ to item $i$ based on the conjunctive condensation rule (Maris, 1999), assuming a value of 1 if person $n$ possesses all the attributes required for item $i$ and a value of 0 if the person lacks at least one of the required attributes; mathematically it is expressed as:

$$\eta_{ni} = \prod_{k=1}^{K} w_{nik} = \prod_{k=1}^{K} \alpha_{nk}^{q_{ik}}, \qquad (2)$$

where $w_{nik}$ can be treated as the latent response on item $i$ for person $n$ to attribute $k$.

```
1.  model{
2.    for (n in 1:N) {
3.      for (i in 1:I) {
4.        for (k in 1:K) {w[n, i, k] <- pow(alpha[n, k], Q[i, k])}
5.        eta[n, i] <- prod(w[n, i, 1:K])
6.        p[n, i] <- g[i] + (1 - s[i] - g[i]) * eta[n, i]
7.        Y[n, i] ~ dbern(p[n, i])}
8.      for (k in 1:K) {alpha[n, k] <- all.patterns[c[n], k]}
9.      c[n] ~ dcat(pai[1:C])}
10.   pai[1:C] ~ ddirch(delta[1:C])
11.   for (i in 1:I) {
12.     s[i] ~ dbeta(a.s, b.s)
13.     g[i] ~ dbeta(a.g, b.g)  T(, 1 - s[i])}}
```
**Table 1**. The DINA model.

Table 1 presents the **JAGS** code to fit the DINA model. The codes are elaborated as follows.

Line 1 signals the starting of the model. Lines 2 to 7 specified the measurement model, lines 8 to 10 are the unstructured latent structural model and priors, and lines 11 to 13 are the priors assumed for the item parameters.

Part of the parameters in Table 1 are assigned with some previously defined values, including `all.patterns`, `C`, `delta`, `Y`, and `Q`. Specifically, `C` is the number of all possible attribute profiles, typically is $2^K$; `all.patterns` is a given $C$-by-$K$ matrix that contains all possible attribute patterns: one for each row; `delta` is the scale parameter vector of the Dirichlet distribution. For generalization, setting `delta` = (1, 1, ..., 1), which means the mixing proportion, `pai`, for all possible patterns follow a non-informative uniform prior distribution; `Y` is a $N$-by-$I$ item response matrix; `Q` is the $I$-by-$K$ Q-matrix. More details about the using of these previously defined parameters in the JAGS code will be illustrated in a subsequent section, titled "An Empirical Example: A Tutorial".

Under the unstructured latent structural model, line 8 describes the method to obtain attributes: $\alpha_{nk} = \alpha_{ck}$, where $c \in \{1, ..., C\}$, indicating person $n$'s attribute profile, is assumed to follow a categorical distribution, with the mixing proportion of the $c$-th pattern.

Since $s_i$ and $g_i$ are on the probability scale, the Beta distributions are often specified as their prior distributions. Lines 12 and 13, as suggested by Culpepper (2015), the scale parameters of the Beta distributions can be assigned as `a.s = b.s = a.g = b.g = 1`, which is identical to a linearly truncated bivariate non-informative uniform prior for $s_i$ and $g_i$ and a monotonicity restriction ($g_i < 1 - s_i$) is specified. In the Bayesian estimation, informative priors might be used according to some previous experiences. For example, according to the results in some previous studies (e.g., Chen, Culpepper, Chen, Douglas, 2018; DeCarlo, 2012; de la Torre & Douglas, 2004; Zhan et al., in press), the quality of items in the fraction subtraction test (Tatsuoka, 1990) are relatively good. In such cases, more informative priors can be used by setting `a.s` = 1, `b.s` = 3, `a.g` = 1, and `b.g` = 3.

In addition, when no prior information is available for the scale parameters a prior on the scale parameters, which is called a hyperprior, can be used. Some extra lines can be added as follows,

```
a.s ~ dunif (0.1, 5)
b.s ~ dunif (0.1, 5)
a.g ~ dunif (0.1, 5)
b.g ~ dunif (0.1, 5)
```

Then, `a.s`, `b.s`, `a.g`, and `b.g` can be estimated.

**JAGS** code of the reparameterized DINA (RDINA) model (DeCarlo, 2011) model is presented in Table 2. The RDINA model is equivalent to the regular DINA model, which can be treated as a restricted version of the LCDM. Because, the RDINA model only allows the $K$-way interaction effect parameter to be freely estimated (see Equation 8). The RDINA model can be expressed as:

$$p_{ni} = P(Y_{ni} = 1 \mid \boldsymbol{\alpha}_n) = \frac{\exp(\lambda_{0,i} + \lambda_{(K),i}\eta_{ni})}{1 + \exp(\lambda_{0,i} + \lambda_{(K),i}\eta_{ni})}, \tag{3}$$

where the intercept parameter ($\lambda_{0,i}$) defines the log-odds of a correct response to item $i$ for a person who is not a master of either one of the attributes; $\lambda_{(K),i}$ is the $K$-way interaction effect parameter for item $i$. In this formulation, the regular $g_i$ and $s_i$ parameters in Equation 1 can be described as:

$$p_{ni} = \begin{cases} \dfrac{\exp(\lambda_{0,i})}{1 + \exp(\lambda_{0,i})} = g_i & \text{if } \eta_{ni} = 0 \\[3mm] \dfrac{\exp(\lambda_{0,i} + \lambda_{(K),i})}{1 + \exp(\lambda_{0,i} + \lambda_{(K),i})} = 1 - s_i & \text{if } \eta_{ni} = 1 \end{cases}. \tag{4}$$

```
1.  model{
2.    for (n in 1:N){
3.      for (i in 1:I){
4.        for (k in 1:K){w[n, i, k] <- pow(alpha[n, k], q[i, k])}
5.        eta[n, i] <- prod(w[n, i, 1:K])
6.        logit(p[n, i]) <- lamda0[i] + lamdaK[i] * eta[n,i]
7.        Y[n,i] ~ dbern(p[n, i])}}
8.    for (k in 1:K) {alpha[n, k] <- all.patterns[c[n], k]}
9.    c[n] ~ dcat(pai[1:C])}
10.   pai[1:C] ~ ddirch(delta[1:C])
11.   for(i in 1:I){
12.     lamda0[i]~dnorm(mean.lamda0, pr.lamda0)
13.     lamdaK[i]~dnorm(mean.lamdaK, pr.lamdaK) T(0, )}}
```

**Table 2**. The RDINA model.

Lines 4 to 10 specifies the model. Line 12 specifies the distribution for the $\lambda_{0,i}$ parameter. A normal prior distribution is assumed targeting at a mean of –1.096, i.e., `mean.lamda0 = –1.096`. This is equivalent to a mean guessing value, $g_i$ of 0.25, which equals to the random guessing probability of a four-option item. Line 13 specifies the distribution for the $\lambda_{(K),i}$ parameter. A normal prior distribution is assumed targeting at a mean of 2.192, i.e., `mean.lamdaK = 2.192`. This makes the mean value of $s_i$ also equals to 0.25. For generalization, less informative prior is assumed. Then, the variances of prior distributions for $\lambda_{0,i}$ and $\lambda_{(K),i}$ parameters can be set at 4. **JAGS** parameterizes the normal distribution in terms of precision (i.e., the inverse of the variance). Thus variance of 4 needs to be converted to a precision of `pr.lamda0 = 0.25` and `pr.lamdaK = 0.25` in lines 12 and 13, respectively. In addition, hyperpriors also can be used here, such as

$$\text{mean.lamda0} \sim \text{dnorm}(-1.096, 0.5)$$
$$\text{pr.lamda0} \sim \text{dgamma}^3(1, 1)$$

Further, the monotonicity restriction ($g_i < 1 - s_i$) is realized by constraining $\lambda_{(K),i}$ parameters to be positive. Thus, a truncated normal distribution is specified for $\lambda_{(K),i}$ in Line 13 by truncation `T(0, )`.

---

[3] Description about the gamma distribution can be found in the comments on Table 9.

## The DINO Model

The DINO model, similar to the DINA model, models the probability of a correct response as a function of a slipping parameter, $s_i$, and a guessing parameter, $g_i$. However, the ideal response, $\eta_{ni}$, in the DINO model is modeled based on the disjunctive condensation rule (Maris, 1999) rather than the conjunctive condensation rule as in the DINA model. $\eta_{ni}$ is expressed as

$$\eta_{ni} = 1 - \prod_{k=1}^{K} w_{nik} = 1 - \prod_{k=1}^{K} (1 - \alpha_{nk})^{q_{ik}} , \qquad (5)$$

which is an indicator of whether person $n$ has mastered at least one of the required attributes for item $i$. Thus, $\eta_{ni} = 1$ for any person having mastered one or more of the item's required attributes; $\eta_{ni} = 0$ for a person who has mastered none of the required attributes. Although the DINO model shares a dual relationship with the DINA model (Köhn & Chiu, 2016), directly fit the DINO model to the data is easier for practitioners.

Table 3 presents the **JAGS** code for the DINO model. The differences between the DINO model and the DINA model can be handled easily by **JAGS**, as shown in Lines 4 and 5 in Tables 1 and 3, respectively.

```
1.  model{
2.    for (n in 1:N) {
3.      for (i in 1:I) {
4.        for (k in 1:K) {w[n, i, k] <- pow(1 - alpha[n, k], Q[i, k])}
5.        eta[n, i] <- 1 - prod(w[n, i, 1:K])
6.        p[n, i] <- g[i] + (1 - s[i] - g[i]) * eta[n, i]
7.        Y[n, i] ~ dbern(p[n, i])}
8.      for (k in 1:K) {alpha[n, k] <- all.patterns[c[n], k]}
9.      c[n] ~ dcat(pai[1:C])}
10.   pai[1:C] ~ ddirch(delta[1:C])
11.   for (i in 1:I) {
12.     s[i] ~ dbeta(1, 1)
13.     g[i] ~ dbeta(1, 1)  T(, 1 - s[i])}}
```
**Table 3**. The DINO model.

## The LLM

The LLM (also called the compensatory reparameterized unified model, C-RUM) is constructed based on the compensatory condensation rule (Maris, 1999). The LLM can be expressed as:

$$p_{ni} = P(Y_{ni} = 1 \mid \boldsymbol{\alpha}_n) = \frac{\exp(\lambda_{0,i} + \sum_{k=1}^{K} \lambda_{k,i} w_{nik})}{1 + \exp(\lambda_{0,i} + \sum_{k=1}^{K} \lambda_{k,i} w_{nik})} = \frac{\exp(\lambda_{0,i} + \sum_{k=1}^{K} \lambda_{k,i} \alpha_{nk} q_{ik})}{1 + \exp(\lambda_{0,i} + \sum_{k=1}^{K} \lambda_{k,i} \alpha_{nk} q_{ik})}, \qquad (6)$$

where $\lambda_{k,i}$ is the $k$-th main effect parameter and all $\lambda_{k,i} \geq 0$. In the LLM, the lowest correct response probability is $\frac{\exp(\lambda_{0,i})}{1+\exp(\lambda_{0,i})}$ denotes the probability of a correct response to item $i$ without mastering any of the required attributes. The probability is then increased as a function of each required attribute that is mastered, as defined by $\lambda_{k,i}$. Finally, the highest probability is $\frac{\exp(\lambda_{0,i} + \sum_{k=1}^{K} \lambda_{k,i} q_{ik})}{1+\exp(\lambda_{0,i} + \sum_{k=1}^{K} \lambda_{k,i} q_{ik})}$ denotes the probability of an incorrect response to item $i$ with mastering all the required attributes.

```
1.  model{
2.    for (n in 1:N){
3.      for (i in 1:I){
4.        for (k in 1:K){w[n, i, k] <- alpha[n, k] * Q[i, k]}
5.        eta[n, i] <- inprod(lamda[i, 1:K], w[n, i, 1:K])
6.        logit(p[n, i]) <- lamda0[i] + eta[n, i]
7.        Y[n, i] ~ dbern(p[n, i])}
8.      for (k in 1:K) {alpha[n, k] <- all.patterns[c[n], k]}
9.      c[n] ~ dcat(pai[1:C])}
10.   pai[1:C] ~ ddirch(delta[1:C])
11.   for(i in 1:I){
12.     lamda0[i] ~ dnorm(-1.096, 0.25)
13.     for(k in 1:K){
14.       lamda[i, k] <- xlamda[i, k] * Q[i, k]
15.       xlamda[i, k] ~ dnorm(0, 0.25) T(0, )}}}
```

**Table 4.** The LLM.

**JAGS** code of the LLM was presented in Table 4. For one item, the number of the main effect parameters is $\sum_{k=1}^{K} q_{ik}$, which is the number of attributes assessed by this item. For example, if an item requires the first two attributes and the test requires three attributes in total, the number of main effect parameters in this item is two rather than three. Thus, only two main effect parameters need be monitored and reported. A prior on the main effect parameters can be induced by defining auxiliary parameters `xlamda` that come from the truncated normal distribution. `lamda` was used for monitoring and final reporting.

## The rRUM

In the DINA model, the aberrant responses are modeled at the item-level. However, in practice, it may seem reasonable that a respondent lacking only one of the measured attributes has a higher chance of a correct response than a respondent who has not mastered any of the measured attributes. To further differentiate between respondents who have not mastered at least one attribute, the noisy-inputs, deterministic "and" gate (NIDA) model (e.g., Junker & Sijtsma, 2001) models the aberrant responses at the attribute-level but with equal constraints across items. A straightforward extension of the NIDA model is the generalized NIDA (G-NIDA) model (de la Torre, 2011) where the slipping and guessing parameters are allowed to vary across items. Thus, there are $2\sum_{i=1}^{I}\sum_{k=1}^{K}q_{ik}$ parameters to be estimated, which makes the model unidentifiable (Jiang, 1996; Culpepper & Hudson, 2017). To make the G-NIDA model identified, Hartz (2002) proposed the rRUM which is a reparameterized version of the G-NIDA model (Culpepper & Hudson, 2017; de la Torre, 2011). The rRUM can be expressed as:

$$p_{ni} = P(Y_{ni} = 1 \mid \boldsymbol{\alpha}_n) = \pi_i^* \prod_{k=1}^{K} r_{ik}^{*w_{nk}} = \pi_i^* \prod_{k=1}^{K} r_{ik}^{*(1-\alpha_{nk})q_{ik}} \ , \tag{7}$$

where $\pi_i^*$ is the baseline parameter that defines the probability of a correct response to item $i$ given that all required attributes; $r_{ik}^*$ is the penalty parameter for not having mastered a required attribute $k$. In the rRUM, there are $\sum_{i=1}^{I}(1+\sum_{k=1}^{K}q_{ik})$ parameters to be estimated. The JAGS code for this model is given in Table 5.

```
1. model{
2.   for(n in 1:N){
3.     for(i in 1:I){
4.       for(k in 1:K){w[n,i,k] <- (1 - alpha[n,k])*Q[i, k]}
5.       p[n, i] <- pai_star[i] * prod(pow(r_star[i, 1:K],w[n, i, 1:K]))
6.       Y[n, i] ~ dbern(p[n, i])}
7.     for(k in 1:K) {alpha[n, k] <- all.patterns[c[n], k]}
8.     c[n] ~ dcat(pai[1:C])}
9.   pai[1:C] ~ ddirch(delta[1:C])
10.  for(i in 1:I){
11.    pai_star[i] ~ dbeta(a.pai_star, b.pai_star)
12.    for(k in 1:K){
13.      r_star[i,k] <- xr_star[i, k] * Q[i, k]
13.      xr_star[i,k] ~ dbeta(a.xr_star, b.xr_star)}}}
```

**Table 5.** The rRUM.

Following the same sequence, the model is first specified from Lines 4 to 6, and priors are specified in the following lines. Please note the distinction between `pai` and `pai_star` in Table 5. The former is the mixing proportion for all possible patterns while the latter is the baseline item parameter.

For one item, only $\sum_{k=1}^{K}q_{ik}$ penalty parameters need to be monitored and reported. A prior on the penalty parameters can be induced by defining auxiliary parameters `xr_star` that are assumed from a Beta distribution. `r_star` is utilized for monitoring and final reporting purposes.

The baseline and penalty parameters both are restricted to values between 0 and 1. Thus, the Beta distributions are used as the priors. For non-informative priors, `a.pai_star`, `b.pai_star`, `a.xr_star`, and `b.xr_star` can be set as 1. In contrast, according to the meaning of these two parameters, more informative priors can be set as `a.pai_star` = 3, `b.pai_star` = 1, `a.xr_star` = 3, and `b.xr_star` = 1.

## The LCDM

Among the CDMs, the LCDM is general enough to encompass many popular CDMs (e.g., the DINA model, the DINO model, the rRUM, and the LLM), which are the special cases by imposing different constraints to the item parameters (Chiu & Köhn, 2016; Henson et al., 2009; Rupp et al., 2010). In the LCDM, the correct response probability for person $n$ on item $i$ is defined as follows:

$$p_{ni} = P(Y_{ni} = 1 \mid \boldsymbol{\alpha}_n) = \frac{\exp(\lambda_{0,i} + \sum_{k=1}^{K} \lambda_{k,i} \alpha_{nk} q_{ik} + \sum_{k=1}^{K-1} \sum_{k'=k+1}^{K} \lambda_{kk',i} \alpha_{nk} \alpha_{nk'} q_{ik} q_{ik'} + \dots + \lambda_{(K),i} \prod_{k=1}^{K} \alpha_{nk} q_{ik})}{1 + \exp(\lambda_{0,i} + \sum_{k=1}^{K} \lambda_{k,i} \alpha_{nk} q_{ik} + \sum_{k=1}^{K-1} \sum_{k'=k+1}^{K} \lambda_{kk',i} \alpha_{nk} \alpha_{nk'} q_{ik} q_{ik'} + \dots + \lambda_{(K),i} \prod_{k=1}^{K} \alpha_{nk} q_{ik})}, (8)$$

where the intercept parameter, $\lambda_{0,i}$, defines the log-odds of a correct response for a person who does not master any attribute; $\lambda_{k,i}$ is the main effect of $\alpha_{nk}$; $\lambda_{kk',i}$ is the two-way interaction effect of $\alpha_{nk}$ and $\alpha_{nk'}$; $\lambda_{(K),i}$ is the $K$-way interaction effect. To keep $p_{ni}$ increase as the number of mastered attributes increase, $\lambda_{k,i}$s and all interaction effects are typically non-negative. The interaction effects can take on any values.

For simplicity, we assume that only three attributes are required by a test, which means there are three main effects, three two-way interaction effects, and one three-way interaction in the LCDM. Then the corresponding **JAGS** code is presented in Table 6.

In the LCDM, the number of the main effect parameters is $\sum_{k=1}^{K} q_{ik}$ ; the number of interaction effect parameters is limited by the highest number of required attributes in the Q-matrix. For example, if one test requires five attributes but no item simultaneously require more than three attributes. Then the highest-way interaction in the LCDM is three rather than five. Similar to the LLM, the priors on the item parameters can be induced by defining auxiliary parameters (e.g., xlamda1).

By setting different constraints, the LCDM can be transferred into different CDMs. For example, if we set all interaction effect parameters in lines 25 to 28 to zeros, then the code in Table 6 is equivalent to the code in Table 4, namely, the LCDM reduced to the LLM:

```
lamda12[i] <- 0
lamda13[i] <- 0
lamda23[i] <- 0
lamda123[i] <- 0
```

```
1.  model{
2.  for(n in 1:N){
3.    for(i in 1:I){
4.      for (k in 1:K){w[n, i, k] <- alpha[n, k] * Q[i, k]}
5.  eta1[n, i] <- lamda1[i] * w[n, i, 1] + lamda2[i] * w[n, i, 2] + lamda3[i] * w[n,
    i, 3]
6.  eta2[n, i] <- lamda12[i] * w[n, i, 1] * w[n, i, 2] + lamda13[i] * w[n, i, 1]
    * w[n, i, 3] + lamda23[i] * w[n, i, 2] * w[n, i, 3]
7.  eta3[n, i] <- lamda123[i] * w[n, i, 1] * w[n, i, 2] * w[n, i, 3]
8.      logit(p[n, i]) <- lamda0[i] + eta1[n, i] + eta2[n, i] + eta3[n, i]
9.      Y[n, i] ~ dbern(p[n, i])}
10.   for(k in 1:K) {alpha[n, k] <- all.patterns[c[n], k]}
11.   c[n] ~ dcat(pai[1:C])}
12.   pai[1:C] ~ ddirch(delta[1:C])
13.   for(i in 1:I) {
14.     lamda0[i] ~ dnorm(-1.096, 0.25)
15.     xlamda1[i] ~ dnorm(0, 0.25) T(0, )
16.     xlamda2[i] ~ dnorm(0, 0.25) T(0, )
17.     xlamda3[i] ~ dnorm(0, 0.25) T(0, )
18.     xlamda12[i] ~ dnorm(0, 0.25)
19.     xlamda13[i] ~ dnorm(0, 0.25)
20.     xlamda23[i] ~ dnorm(0, 0.25)
21.     xlamda123[i] ~ dnorm(0, 0.25)
22.     lamda1[i] <- xlamda1[i] * Q[i, 1]
23.     lamda2[i] <- xlamda2[i] * Q[i, 2]
24.     lamda3[i] <- xlamda3[i] * Q[i, 3]
25.     lamda12[i] <- xlamda12[i] * Q[i, 1] * Q[i, 2]
26.     lamda13[i] <- xlamda13[i] * Q[i, 1] * Q[i, 3]
27.     lamda23[i] <- xlamda23[i] * Q[i, 2] * Q[i, 3]
28.     lamda123[i] <- xlamda123[i] * Q[i, 1] * Q[i, 2] * Q[i, 3]}}
```

**Table 6**. The LCDM.

## The Higher-Order Latent Structural Model

In CDMs, the number of all possible attribute patterns is typically $2^K$. The unstructured latent structural model that was used in previous sections requires $2^K - 1$ structural parameters for such $2^K$ possible patterns, and lead to a substantial computational burden when there are many attributes. de la Torre and Douglas (2004) proposed a solution to reduce the calculations connected with the estimation of CDM parameters by involving a higher-order latent structure beyond the attributes. They proposed a higher-order latent structural model where all attributes are assumed to be conditionally independent given a continuous latent trait θ:

$$p_{nk} = P(\alpha_{nk} \mid \theta_n) = \frac{\exp(\xi_k \theta_n - \beta_k)}{1 + \exp(\xi_k \theta_n - \beta_k)} \quad , \qquad (9)$$

where $p_{nk}$ is the probability of person $n$ mastering attribute $k$ given θ. $\beta_k$ and $\xi_k$ denote the intercept and slope parameter of the $k$-th attribute, respectively, and θ is assumed as $N(0, 1)$ for model identification. Owing to this higher-order structure, the number of attribute parameters to be estimated is only $2K$ (i.e., $K$ attribute intercept parameters and $K$ attribute slope parameters) rather than $2^K - 1$. Because the number of parameters grows linearly, not exponentially, this formulation significantly reduces the computational burden. Theoretically speaking, as a latent structural model, Equation 9 can be employed in any CDM. For simplicity, the DINA model is used to illustrate how to incorporate the higher-order latent structural model into the DINA model to yield the HO-DINA model.

```
1.  model{
2.    for (n in 1:N) {
3.      for (i in 1:I) {
4.        for (k in 1:K) {w[n, i, k] <- pow(alpha[n, k], Q[i, k])}
5.          eta[n, i] <- prod(w[n, i, 1:K])
6.          p[n, i] <- g[i] + (1 - s[i] - g[i]) * eta[n, i]
7.          Y[n, i] ~ dbern(p[n, i])}}
8.    for(n in 1:N){
9.      for(k in 1:K){
10.       logit(prob.a[n, k]) <- xi[k] * theta[n] - beta[k]
11.       alpha[n, k] ~ dbern(prob.a[n, k])}
12.     theta[n] ~ dnorm(0, 1)}
13.   for(k in 1:K){
14.     beta[k] ~ dnorm(mean.beta, pr.beta)
15.     xi[k] ~ dnorm(mean.xi, pr.xi) T(0, )}
16.   for (i in 1:I) {
17.     s[i] ~ dbeta(1, 1)
18.     g[i] ~ dbeta(1, 1)  T(, 1 - s[i])}}
```

**Table 7**. The HO-DINA model.

We use the code from lines 8 to 12 to describe the method to obtain attributes from the higher-order latent structural model. Lines 13 to 15 are the priors of the latent structural parameters. According to the estimated results in previous studies (e.g., Zhan, et al., 2018), the absolute values of $\beta_k$ and $\xi_k$ may be large to a value of 3 or even 4. Thus, the scale parameters are suggested to be set as `mean.beta = 0`, `mean.xi = 0`, `pr.beta = 0.25`, and `pr.xi = 0.25`. Assuming higher $\theta$ values could lead to higher $p_{nk}$, which is not strictly necessary (e.g., if one attribute is a misconception rather than a skill; see Bradshaw & Templin, 2014), we could still restrict $\xi_k > 0$. Specifically, a truncated normal distribution is specified for `xi[k]` in Line 15 by using the `T(0, )` operator.

From the seven examples presented in Tables 1 to 7, an obvious advantage of using **JAGS** is that previous introduced models could be extended easily by altering a few lines of **JAGS** code. In the next three sections, we will further extend CDMs to address the polytomous attribute, the testlet effect, and the longitudinal data.

### A DINA model for Polytomous Attributes

All the models presented are limited to the binary attributes (i.e., mastery or non-mastery). Such binary classification may be difficult to differentiate persons within the same category who master a specific attribute differently. For instance, an extreme case would be a person fully masters an attribute while another person is a borderline master who is slightly above the threshold. Thus, the polytomous attributes and the polytomous Q-matrix (Karelitz 2004; von Davier, 2008) could be a better option for ……. While a binary attribute is related to two mastery status, a polytomous attribute is related to more than two categories (e.g., 0, 1, 2, etc.). This fine-grained sizing helps to provide a more informative diagnosis of respondents. Substantively, the polytomous categories can be different for each attribute with well-defined meanings by content experts. Currently, the ordered category attribute coding (OCAC) framework (Karelitz, 2004) is used to address polytomous attributes (e.g., Chen & de la Torre 2013; Zhan, Bian, & Wang, 2016). In the OCAC framework, the ordinal levels of each attribute are coded as non-negative integers starting from 0, 1 to the highest level. For illustration, the reparameterized polytomous attributes DINA (RPa-DINA) model (Zhan et al., 2016) is demonstrated as an example here. The RPa-DINA model can be expressed as:

$$p_{ni} = P(Y_{ni} = 1 \mid \boldsymbol{\alpha}_n) = g_i + (1 - s_i - g_i)\eta_{ni},$$

$$\eta_{ni} = \prod_{k=1}^{K} w_{nik} = \prod_{k=1}^{K} A_{nik}^{q_{ik}^*}, \tag{10}$$

where $\alpha_{nk}$ is a polytomous variable for person $n$ on attribute $k$, $\alpha_{nk} = l - 1$ if person $n$ masters the $l$-th level ($l = 1, ..., L_k$) of attribute $k$, and let $L_k$ be the number of ordinal levels of attribute $k$. As the first level of attribute $k$ is labeled as 0, $\alpha_{nk} = l - 1$; The polytomous Q-matrix is an $I$-by-$K$ matrix with element $q_{ik} = l - 1$ indicating the $l$-th level of attribute $k$ is required to answer item $i$ correctly; $A_{nik} = I\{\alpha_{nk} \geq q_{ik}\}$ is the ideal response to item $i$ for person $n$ on attribute $k$, where $I\{\cdot\}$ is an indicator function. Thus, $A_{nik} = 1$ if person $n$'s attribute mastery level is at or above the specific attribute level that is required by item $i$, and 0 otherwise; $q_{ik}^* = I\{q_{ik} > 0\}$ indicates whether attribute $k$ is required by item $i$. Typically, the number of possible polytomous attribute patterns is $\prod_{k=1}^{K}(L_k + 1)$. **JAGS** code for the RPa-DINA model is presented in Table 8.

In line 5, `step(x)` equals to 1 if $x \geq 0$, and 0 otherwise. In line 13, `Q_star` is the $I$-by-$K$ binary Q-matrix that was reduced from the polytomous Q-matrix by using $q_{ik}^* = I\{q_{ik} > 0\}$. When $L_k = 2$ for all attributes, the RPa-DINA model is equivalent to the DINA model for binary attributes (see Equations 1 and 2). Therefore, the code in Table 8 can be used directly to describe the DINA model for binary attributes without any modifications.

```
1.  model{
2.   for(n in 1:N){
3.    for(i in 1:I){
4.     for(k in 1:K){
5.     A[n,i,k] <- step(alpha[n, k] - Q[i, k])
6.     w[n,i,k] <- pow(A[n, i, k], Q_star[i, k])}
7.     eta[n,i] <- prod(w[n, i, 1:K])
8.     p[n,i] <- g[i] + (1 - s[i] - g[i]) * eta[n, i]
9.     Y[n,i] ~ dbern(p[n, i])}
10.  for(k in 1:K){alpha[n, k] <- all.patterns[c[n], k]}
11.  c[n]~dcat(pai[1:C])}
12.  pai[1:C] ~ ddirch(delta[1:C])
13.  for(i in 1:I){for(k in 1:K){Q_star[i, k] <- step(Q[i, k] - 1)}}
14.  for(i in 1:I){
15.   s[i]~dbeta(1, 1)
16.   g[i]~dbeta(1, 1) T(,1 - s[i])}}}
```

**Table 8.** The RPa-DINA model.

Note that $q_{ik}^*$ is useless for the conjunctive condensation rule (e.g., the DINA model), because $\prod_{k=1}^{K} A_{nik}^{q_{ik}^*} = \prod_{k=1}^{K} A_{nik}$; however, it is necessary for the disjunctive condensation rule such as in the DINO model, $\eta_{ni} = 1 - \prod_{k=1}^{K} w_{nik} = 1 - \prod_{k=1}^{K}(1 - A_{nk})^{q_{ik}^*}$ and the compensatory condensation rule such as in the LLM , $\eta_{ni} = \sum_{k=1}^{K} w_{nik} = \sum_{k=1}^{K} A_{nk} q_{ik}^*$.

## A DINA Model for Testlet Design

Testlets have been widely adopted in educational and psychological tests. A testlet is a cluster of items that share a common stimulus (Wainer & Kiely, 1987). For example, in a reading comprehension test, a testlet is formed as a bundle of items based on one reading passage. Local item dependence among items within a testlet is called as the testlet effect. The testlet effect could be an indication of a noise dimension. In the IRT framework, testlet effects are accounted for by adding a set of additional random effect parameters to standard IRT models: one for each testlet (Wainer et al. 2007) or multiples for each testlet (Zhan, Wang, Wang, & Li, 2014). In practice, testlets can be used in cognitive diagnosis assessment. Although it is not conceptually challenging to add a set of random effect parameters into CDMs, limited efforts have been made to the development of testlet CDMs (e.g., Hansen et al., 2016; Liao & Jiao, 2016; Zhan, Li, Wang, Bian, & Wang, 2015; Zhan, Liao, & Bian, 2018).

For illustration, the RDINA model (see Table 2) is used as a template, and this method can be extended easily to the LCDM and other cases. To address the testlet effect, a random effect parameter, $\gamma_{nd(i)}$, is added to the RDINA model:

$$p_{ni} = P(Y_{ni} = 1 \mid \boldsymbol{\alpha}_n, \gamma_{nd(i)}) = \frac{\exp(\lambda_{0,i} + \lambda_{(K),i}\eta_{ni} + \gamma_{nd(i)})}{1 + \exp(\lambda_{0,i} + \lambda_{(K),i}\eta_{ni} + \gamma_{nd(i)})}, \qquad (11)$$

where the random effect parameter, $\gamma_{nd(i)}$, is assumed from a normal distribution $\gamma_{nd(i)} \sim N(0, \sigma^2_{\gamma_d})$, and $\sigma^2_{\gamma_d}$ indicates the magnitude of the testlet effect for testlet $d$. Other model parameters remain the same as the models illustrated above.

```
1.  model{
2.   for(n in 1:N){
3.    for(i in 1:I){
4.     for(k in 1:K){w[n, i, k] <- pow(alpha[n, k], Q[i, k])}
5.     eta[n, i] <- prod(w[n, i, 1:K])
6.     logit(p[n, i]) <- lamda0[i] + lamdaK[i] * eta[n,i] + gamma[n, d[i]]
7.     Y[n, i] ~ dbern(p[n, i])}
8.    for(k in 1:K) {alpha[n, k] <- all.patterns[c[n], k]}
9.    c[n] ~ dcat(pai[1:C])}
10.  pai[1:C] ~ ddirch(delta[1:C])
11.  for(n in 1:N){
12.   for(m in 1:M){gamma[n,m] ~ dnorm(0, pr_gamma[m])}
13.   gamma[n,M+1]<-0}
14.  for(m in 1:M){
15.   pr_gamma[m] ~ dgamma(1,1)
16.   Sigma_gamma[m] <- 1 / pr_gamma[m]}
17.  for(i in 1:I){
18.   lamda0[i]~dnorm(-1.096, 0.25)
19.   lamdaK[i]~dnorm(0, 0.25) T(0, )}}
```

**Table 9**. The Testlet-DINA Model.

The number of testlets (i.e., M) needs to be specified, so does the testlet identifier vector $d$. The element in vector $d$ (i.e., d[i]) is used to indicate the testlet that item $i$ is associated with. It should be noted that, if item $i$ is a standalone item, gamma[i, M+1] is set to be 0 as in in line 13. For example, a test consists of 10 items, 2 testlets with 4 items associated with each testlet, and last two items are standalone items. Vector *d* should be set as

```
d = c (1, 1, 1, 1, 2, 2, 2, 2, 3, 3).
```

In this model, `Sigma_gamma` is the variance of testlet effect, $\sigma_{\gamma_d}^2$, to be monitored and estimated in line 16. **JAGS** parameterizes the normal distribution regarding precision-the inverse of the variance. The **JAGS** can not specify an inverse-gamma distribution. Typically, a gamma prior on the inverse of the monitored parameter is specified. Thus, lines 15 to 16 specifies an inverse-gamma prior on the `Sigma_gamma[m]` parameter.

In addition to the unstructured latent structural model, the higher-order latent structural model (Equation 9) also can be introduced, see the next section.

## A DINA Model for Longitudinal Data

Providing diagnostic feedback about growth is crucial to formative decisions such as targeted remedial instructions or interventions. Measuring individual growth or change relies on longitudinal data collected over multiple measures of achievement construct along the growth trajectory. However, few studies focus on measuring growth in terms of several related attributes over multiple occasions (e.g., Li et al., 2016; Wang, Yang, Culpepper, & Douglas, 2018; Zhan, Jiao, & Liao, 2017). Unlike continuous latent traits in IRT models, the attributes in CDMs are categorical. Therefore, the methods for modeling growth in the IRT framework may not be directly extended to capture growth in the mastery of attributes.

Currently, there are two main approaches to analyzing longitudinal data in cognitive diagnosis. The first is from the latent class modeling perspective (Chen, Culpepper, Wang, & Douglas, 2018; Li et al., 2016; Wang et al., 2018), which can all be taken as a particular case or an application of the mixture hidden Markov model (Vermunt, Tran, & Magidson, 2008). The second is from the IRT modeling perspective, such as the longitudinal higher-order DINA (Long-DINA) model (Zhan et al., 2017), which uses the variance-covariance-based method by assuming multiple continuous higher-order latent traits (see Equation 9) follow a multivariate normal distribution.

As potential local item dependence among anchor (or repeated) items can be taken into account, and also following with the description of the testlet-DINA model in Table 9, the Long-DINA model is introduced in this paper. Essentially, the Long-DINA model can be taken as an extension of the testlet-DINA model by incorporating a multidimensional higher-order latent structure to take into account the correlations among multiple latent attributes that are examined across different occasions. The Long-DINA model can be expressed as:

$$\text{First-level:} \quad p_{nit} = P(Y_{nit} = 1 \mid \boldsymbol{\alpha}_{nt}, \gamma_{nd(i)}) = \frac{\exp(\lambda_{0,i,t} + \lambda_{(K),i,t} \prod_{k=1}^{K} \alpha_{nkt}^{q_{ikt}} + \gamma_{nd(i)})}{1 + \exp(\lambda_{0,i,t} + \lambda_{(K),i,t} \prod_{k=1}^{K} \alpha_{nkt}^{q_{ikt}} + \gamma_{nd(i)})},$$

$$\text{Second-level:} \quad p_{nkt} = P(\alpha_{nkt} \mid \theta_{nt}) = \frac{\exp(\xi_k \theta_{nt} - \beta_k)}{1 + \exp(\xi_k \theta_{nt} - \beta_k)}, \ \boldsymbol{\theta}_n = (\theta_{n1}, \ldots, \theta_{nT})',$$

$$\text{Third-level:} \quad \boldsymbol{\theta}_n = (\theta_{n1}, \ldots, \theta_{nT})' \sim \text{MVN}_T(\boldsymbol{\mu}_\theta, \boldsymbol{\Sigma}_\theta), \tag{12}$$

where $Y_{nit}$ denotes the response of person $n$ to item $i$ on occasion $t$; $\boldsymbol{\alpha}_{nt} = (\alpha_{n1t}, \ldots, \alpha_{nKt})'$ denotes person $n$'s attribute profile on occasion $t$; $\lambda_{0,i,t}$ and $\lambda_{(K),i,t}$ are the intercept and $K$-way interaction effect parameter for item $i$ on occasion $t$, respectively; $q_{ikt}$ is the element in the Q-matrix on occasion $t$; $\gamma_{nd(i)} \sim N(0, \sigma_{\gamma_d}^2)$ is the specific dimension parameter for person $n$, used to account for local item dependence among anchor (or repeated) items on different occasions; $\theta_{nt}$ is person $n$'s general ability on occasion $t$, $\xi_{kt}$ and $\beta_{kt}$ are the slope and intercept parameters of attribute $k$ on occasion $t$, respectively. The same attributes and the same underlying latent construct are assumed to be measured on different occasions (Bianconcini, 2012), i.e., $K_t = K$. Thus, the slope and intercept parameters of the $k^{\text{th}}$ attribute are constrained to be constants across occasions, i.e., $\xi_{kt} = \xi_k$ and $\beta_{kt} = \beta_k$. $\theta_n s$ are assumed to be independent of $\gamma_n s$. $\boldsymbol{\mu}_\theta = (\mu_1, ..., \mu_T)'$ is the mean vector of multidimensional higher-order latent traits and $\boldsymbol{\Sigma}_\theta$ is a variance and covariance matrix. As a starting and a reference point for subsequent occasions, $\theta_{n1}$ is constrained to follow a standard normal distribution.

For illustration, we assume two occasions ($T = 2$), 10 items on each occasion, and the first 2 items ($M = 2$) on each occasion are used as anchor items. Then the corresponding **JAGS** code is given in Table 10.

For the test scenario specified above, `I` and `d[i]` should be set as `I = c(10, 10)` and `d = c (1, 2, 3, 3, 3, 3, 3, 3, 3, 3, 1, 2, 3, 3, 3, 3, 3, 3, 3, 3).` Lines 23 to 25 are prior distributions for items on the first occasion; lines 26 to 29 are used for anchor items; lines 30 to 32 are prior distributions for non-anchor items on the second occasion.

The multivariate normal distribution in **JAGS** is also parameterized in the precision matrix, which is the inverse of the covariance matrix. Thus, the covariance matrix `Sigma_theta`, which needs to be monitored and estimated, is inverted on line 42, the resulting precision matrix `pr_theta` is then used in the `dmnorm` function. Typically, the inverse-Wishart distributions are used to specify priors for the covariance matrices. However, an inverse-Wishart prior cannot be used for $\boldsymbol{\Sigma}_\theta$ (i.e., `Sigma_theta`), because the variance of $\theta_{n1}$ is set to 1. To solve this problem, $\boldsymbol{\Sigma}_\theta$ can be reparameterized in terms of its Cholesky decomposition as $\boldsymbol{\Sigma}_\theta = \boldsymbol{\Delta}_\theta \boldsymbol{\Delta}_\theta'$ (Curtis, 2010; Zhan et al., 2018), where $\boldsymbol{\Delta}_\theta = \begin{pmatrix} 1 & 0 \\ \varphi & \psi \end{pmatrix}$ is a lower triangular matrix with positive entries on the diagonal and unrestricted off-diagonal entries , and $\boldsymbol{\Delta}_\theta'$ is the conjugate transpose of $\boldsymbol{\Delta}_\theta$. Therefore, `L_theta[tt, ttt] ~ dnorm(0, 1)` is used for $\varphi \sim N(0, 1)$ in line 39, and

`L_theta[tt, tt] ~ dgamma(1, 1)` is used for $\psi \sim \text{Gamma}(1, 1)$ in line 37.

```
1.  model{
2.    for(t in 1:T){
3.      for(n in 1:N){
4.        for(i in 1:I[t]){
5.          for(k in 1:K){w[n,i,k,t] <- pow(alpha[n, k, t], Q[i, k, t])}
6.          eta[n, i, t] <- prod(w[n, i, 1:K, t])
7.          logit(p[n, i, t]) <- lamda0[i,t]+lamdaK[i,t]*eta[n,i,t]+gamma[n,d[i]]
8.          Y[n, i, t] ~ dbern(p[n, i, t])}}
9.      for(n in 1:N){
10.       for(k in 1:K){
11.         logit(prob.a[n, k, t]) <- xi[k] * theta[n,t] - beta[k]
12.         alpha[n, k, t] ~ dbern(prob.a[n, k, t])}}}
13.     for(n in 1:N){theta[n,1:T] ~ dmnorm(mu_theta[1:T], pr_theta[1:T, 1:T])}
14.     for(k in 1:K){
15.       beta[k] ~ dnorm(0, 0.25)
16.       xi[k] ~ dnorm(0, 0.25) T(0, )
17.     for(n in 1:N){
18.       for(m in 1:M){gamma[n,m] ~ dnorm(0, pr_gamma[m])}
19.       gamma[n, M+1] <- 0}
20.     for(m in 1:M){
21.       pr_gamma[m] ~ dgamma(1, 1)
22.       Sigma_gamma[m] <- 1 / pr_gamma[m]}
23.     for(i in 1:I[1]){
24.       lamda0[i,1] ~ dnorm(-1.096, 0.25)
25.       lamdaK[i,1] ~ dnorm(0, 0.25) T(0, )}
26.     lamda0[1, 2] <- lamda0[1, 1]
27.     lamda0[2, 2] <- lamda0[2, 1]
28.     lamdaK[1, 2] <- lamdaK[1, 1]
29.     lamdaK[2, 2] <- lamdaK[2, 1]
30.     for(i in 3:I[2]){
31.       lamda0[i,2] ~ dnorm(-1.096, 0.25)
32.       lamdaK[i,2] ~ dnorm(0, 0.25) T(0, )}
33.     mu_theta[1] <- 0
34.     for (t in 2:T){mu_theta[t] ~ dnorm(0, 0.5)}
35.     L_theta[1, 1] <- 1
36.     for(tt in 2:T){
37.       L_theta[tt, tt] ~ dgamma(1, 1)
38.       for(ttt in 1:(tt-1)){
39.         L_theta[tt, ttt] ~ dnorm(0, 1)
40.         L_theta[ttt, tt] <- 0}}
41.     Sigma_theta <- L_theta %*% t(L_theta)
42.     pr_theta[1:T, 1:T] <- inverse(Sigma_theta[1:T, 1:T])}}
```
**Table 10**. The Long-DINA Model.

The Expectation-Maximization algorithm via the flexMIRT (version 3.51; Cai, 2017) was used in Zhan et al. (2017)[4]. However, due to the restriction of the flexMIRT, multiple Q-matrices and attribute patterns on different occasions should be combined and rebuilt together in analysis, which may lead to large computing burden. For example, if $T = 4$ and $K = 5$, then $2^{TK} = 1048576$ attribute patterns need to be estimated in the flexMIRT. By contrast, due to the flexibility of **JAGS**, multiple Q-matrices and attribute patterns on different occasions are used separately (e.g., `Q[i, k, t]` in Table 10), thus, only $2^K \times T = 128$ attribute patterns need to be estimated.

---

[4] To our knowledge, currently, among existing stand-alone software and packages, only the flexMIRT can be utilized to fit the Long-DINA model.

## An Empirical Example: A Tutorial

To demonstrate how to use the **JAGS** code presented in the earlier sections to analyze a real dataset, a fraction subtraction data from de la Torre (2009), originally used by Tatsuoka (1990), was analyzed. The dataset contained a total of 536 people responding to 15 items measuring 5 required attributes. The total number of possible attribute profiles was 32. The Q-matrix can be found in de la Torre (2009). The response data and the Q-matrix can be read in **R** first:

```
setwd("C:/...") #Set working directory
set.seed(12345)
library(CDM) #CDM package is only used to read the fraction subtraction data.
data(data.fraction1) #Read the fraction subtraction data and Q-matrix.
Y <- data.matrix(data.fraction1$data); View(Y)
Q <- data.matrix(data.fraction1$q.matrix); View(Q)
```

This section illustrates how to employ the **R2jags** package and **JAGS** code to analyze the fraction subtraction data step by step. The DINA model and the rRUM were employed and compared. For simplicity, only the DINA model is presented for illustration. Readers can directly adapt the code in **R** for other models.

*Step 1: Construct `all.patterns`*

Within the code, `all.patterns` is a matrix that contains all possible attribute patterns. The `gapp` (generate all possible patterns) function[5] below can be used to help readers to quickly generate `all.patterns` based on your own Q-matrix.

```
gapp <- function(q){
```

---

[5] This function can be used for both binary and polytomous attributes.

```
  K <- ncol(q)
  q.entries <- as.list(1:K)
  for(k in 1:K){q.entries[[k]] <- sort(unique(c(0, q[,k])))}
  attr.patt <- as.matrix(expand.grid(q.entries))}
all.patterns <- gapp(Q); View(all.patterns)
```

*Step 2: Load JAGS Code for the DINA Model*

```
DINA <- function(){
  for (n in 1:N) {
    for (i in 1:I) {
      for (k in 1:K) {w[n, i, k] <- pow(alpha[n, k], Q[i, k])}
      eta[n, i] <- prod(w[n, i, 1:K])
      p[n, i] <- pow((1 - s[i]), eta[n, i]) * pow(g[i], (1 - eta[n, i]))
      Y[n, i] ~ dbern(p[n, i])}
    for (k in 1:K) {alpha[n, k] <- all.patterns[c[n], k]}
    c[n] ~ dcat(pai[1:C])}
  pai[1:C] ~ ddirch(delta[1:C])
  for (i in 1:I) {
    s[i] ~ dbeta(1, 1)
    g[i] ~ dbeta(1, 1) %_% T( , 1 - s[i])}
  ##the posterior predictive model checking##
  for (n in 1:N){
    for (i in 1:I){
      teststat[n,i] <- pow(Y[n, i] - p[n, i], 2)/(p[n, i] * (1 - p[n, i]))
      Y_rep[n,i] ~ dbern(p[n,i])
  teststat_rep[n,i] <- pow(Y_rep[n, i] - p[n, i],2)/(p[n, i] * (1 - p[n, i]))}}
  teststatsum <- sum(teststat[1:N, 1:I])
  teststatsum_rep <- sum(teststat_rep[1:N, 1:I])
  ppp <- step(teststatsum_rep - teststatsum)}
```

In **R**, to overcome the incompatibility, the dummy operator `%_%` should be used before `T( ,`
`1 - s[i])`. The dummy operator `%_%` will be removed before the code is saved as a separate
file. In addition, the posterior predictive model checking (PPMC; Gelman, Carlin, Stern, Dunson,
Vehtari, & Rubin, 2014) is used to evaluate the absolute model-data fit. Posterior predictive
probability (ppp) values near 0.5 indicate that there are no systematic differences between the
realized and predictive values, and thus adequate fit of the model to the data. By contrast, ppp
valuew near 0 or 1 (typically ppp value < 0.05 or ppp value > 0.95) suggest inadequate model fit
(Gelman et al., 2014). The sum of the squared Pearson residuals for person *n* and item *i* (Yan,
Mislevy, & Almond, 2003) is used as a discrepancy measure to evaluate the overall fit of the
model as follows.

$$D(Y_{ni}; \boldsymbol{\alpha}_n) = \sum_{n=1}^{N} \sum_{i=1}^{I} \left( \frac{Y_{ni} - p_{ni}}{\sqrt{p_{ni}(1 - p_{ni})}} \right)^2 ,$$

where $p_{ni}$ is defined the same as in Equation 1. Note that, other kinds of discrepancy measures also can be used for different purposes (e.g., Levy & Mislevy, 2016).

*Step 3: Load the R2jags Package and Data*

```
library(R2jags)
N = nrow(Y) #as an exercise, N = 100 can be used to reduce the time cost
I = nrow(Q)
K = ncol(Q)
C = nrow(all.patterns)
delta = rep(1, C)
jags.data = list("N", "I", "K", "Y", "Q", "C", "all.patterns", "delta")
```

*Step 4: Preliminary study for parameter convergence*

```
pre.parameters <- c("s", "g", "pai")
#s: slipping parameter
#g: guessing parameter
#pai: posterior mixing proportion
jags.inits <- NULL #Initial values are not specified.

pre.sim <- jags(data = jags.data, inits = jags.inits, parameters.to.save =
            pre.parameters, model.file = DINA, n.chains = 2, n.iter = 1000,
            n.thin = 1, DIC = TRUE)
R_convergence <- sum(pre.sim$BUGSoutput$summary[ , 8] >= 1.2) == 0; R_convergence
if(R_convergence == 0){pre.sim.c <- autojags(pre.sim, Rhat = 1.2, n.update = 30)
                    pre.sim.c$n.iter}; pre.sim.c$n.iter
if(R_convergence == 1){pre.sim$n.iter}; pre.sim$n.iter
```

A preliminary study was conducted to get a necessary number of iterations to achieve convergence. In Bayesian CDMs, item parameters and mixing proportions are typically checked for convergence (e.g., Culpepper, 2015a; Zhan, Jiao, Liao, & Bian, in press). In the preliminary study, two Markov chains (`n.chains = 2`) were used with `n.iter = 1000` iterations per chain, with the first half of iterations in each chain as burn-in (in default), the thinning interval was set to be `n.thin = 1` (i.e., without thinning)[6]. Finally, the remaining half iterations were used for model parameter inferences. The potential scale reduction factor, $\hat{R}$, as modified by

---

[6] Sometimes, in order to avoid high autocorrelations between the sampling distributions or to take up less space in memory for large-scale data, the thinning interval can be set to be 5 or larger number.

Brooks and Gelman (1998), was computed to assess the convergence of every parameter. Values of $\hat{R}$ less than 1.1 or 1.2 indicate convergence (Brooks & Gelman, 1998; de la Torre & Douglas, 2004). If any parameter estimate does not reach convergence (i.e., R_convergence is FALSE), updating would automatically continue until all values of $\hat{R}$ are less than 1.2. More stringent rules for convergence can be used by setting Rhat = 1.1 or 1.05 in the autojags function. pre.sim$n.iter or pre.sim.c $n.iter is used to show how many iterations are needed to converge. In this example, 1,000 iterations are necessary for convergence based on the rule of $\hat{R} < 1.2$.

*Step 5: Parameter Estimation*

```
jags.parameters <- c("s", "g", "c", "pai", "ppp")
#c: estimated attribute pattern of each respondent
#ppp: posterior predictive probability
jags.inits <- NULL #Initial values are not specified.

time1 = as.POSIXlt(Sys.time())
sim <- jags(data = jags.data, inits = jags.inits, parameters.to.save =
            jags.parameters, model.file = DINA, n.chains = 2, n.iter = 10000,
            n.burnin = 5000, n,thin = 1, DIC = TRUE)
time2 = as.POSIXlt(Sys.time())
use.time = difftime(time2, time1, units="secs")
```

Although the preliminary study indicates a burn-in of 1,000 iterations is adequate, a burn-in of 5,000 iterations was employed in the study to ensure the stability of the results. Two Markov chains were used with 1,0000 iterations per chain and the first 5,000 iterations in each chain excluded as burn-in. The thinning interval was set to be 1. Finally, 10,000 iterations were used for model parameter inferences. use.time function was used to compute the overall running time for parameter estimation[7].

*Step 6: Save Estimated Parameters*

```
sim1 <- sim$BUGSoutput
E.pattern <- cbind(sim1$median$c)
```

---

[7] All runs reported in this article were on an msi GT72VR 6RD DOMINATOR laptop with a 2.6GHz Intel Core i7 6700HQ CPU, 2133MHz 32GB of memory, and 256GB SanDisk z400s Solid State Drive.

```
E.itempar <- cbind(sim1$mean$g, sim1$mean$s, sim1$sd$g, sim1$sd$s)
write.table(E.pattern, "pattern_DINA.txt")
write.table(E.itempar, "itempar_DINA.txt")
write.table(sim1$summary, "summary_DINA.txt")
write.table(sim1$DIC, "DIC_DINA.txt")
write.table(use.time, "time_DINA.txt")
write.table(sim1$mean$deviance, "deviance_DINA.txt")
write.table(sim1$mean$ppp, "ppp_DINA.txt")
```

The file "summary_DINA.txt" presents the summary statistics based on the sampled values of all monitored parameters. Taking the mixing proportions as an example, Table 11 presents the first three estimated mixing proportions, `pai`, of 32 possible patterns. Note that `pai[1]` to `pai[32]` corresponding to the 32 rows in `all.patterns`, respectively. The posterior mean and the standard deviation can be used as the point estimates of the mixing proportions and their standard errors. The values corresponding to the column named 2.5% and the column named 97.5% can be used as the 95% credible interval. The column named `Rhat` lists the $\hat{R}$. The column named `n.eff` listed the effective number of simulation draws, which can be viewed as the effective sample size for a posterior distribution upon which inferences were based. Additionally, the trace plots for the mixing proportions can be requested by inputting `traceplot(sim, varname = "pai")`.

**Table 11**. Sample Output Results of Mixing Proportions for the DINA model.

|         | mean  | sd    | 2.5%  | 25%   | 50%   | 75%   | 97.5% | Rhat  | n.eff |
|---------|-------|-------|-------|-------|-------|-------|-------|-------|-------|
| pai[1]  | 0.018 | 0.016 | 0.000 | 0.006 | 0.013 | 0.024 | 0.060 | 1.004 | 440   |
| pai[2]  | 0.009 | 0.008 | 0.000 | 0.003 | 0.007 | 0.013 | 0.028 | 1.001 | 3200  |
| pai[3]  | 0.017 | 0.015 | 0.000 | 0.006 | 0.013 | 0.025 | 0.058 | 1.004 | 1000  |

The file "ppp_DINA.txt" contains the PPP value of the DINA model to this data. In addition, the file "deviance_DINA.txt" extracts the posterior mean of `deviance` in the MCMC samples, i.e., –2 log likelihood (–2LL), which can be used to compute some relative model-data fit, e.g., AIC (Akaike, 1974) index. The file "DIC_DINA.txt" given the DIC index (Spiegelhalter,

Best, Carlin, & Van der Linde, 2002), where $DIC = \overline{D} + p_e = \overline{D} + \text{var}(D)/2$, namely, the effective

number of parameters ($p_e$) was computed by $p_e = \text{var}(D)/2$ (Gelman, Carlin, Stern, & Rubin,

2003; Su & Yajima, 2015), where $D$ is the deviance, and $\overline{D}$ is the posterior mean of deviance

(i.e., –2LL, the value in the file of "deviance_DINA.txt"). Note that in Bayesian analysis, the

AIC can be defined as $AIC = \overline{D} + p$ (Congdon, 2003), where $p$ is the number of estimated parameters. In addition, as mentioned by Gelman et al. (2014), BIC (Schwarz, 1978) has a different goal from AIC and DIC, namely, "*BIC is not intended to predict out-of-sample model performance but rather is designed for other purposes, we do not consider it further here.*" (p.

175)[8]. In addition, the file "pattern_DINA.txt" is for the estimated attribute patterns. As a

categorical value, the posterior mode of `c` is treated as the estimated value in this study, and the

value of `c`, i.e., 1 to 32, corresponding to the 32 rows in `all.patterns`, respectively. The file

"time_DINA.txt" summarizes the overall computing time.

Table 12 presents the model fit comparison between the DINA model and the rRUM model

to the fraction subtraction data. The rRUM model was identified as a better fitting model based

on AIC, DIC, and a ppp value of 0.701. It took about 1535 and 3370 seconds to run the DINA

model and the rRUM respectively.

Table 13 presents the estimates of item parameters for two models. Table 14 presents the estimates of the mixing proportions for the two models. Noted that the comparison between these two models is beyond the scope of this paper. Thus, no further explanation of the results is provided.

---

[8]    Note that, if the reader still want to report BIC, in Bayesian analysis, the BIC can be defined as $BIC = \overline{D} + (\log N - 1)p$.

**Table 12**. Model Fit and Computing Time in the Empirical Example.

| Model | ppp | NP | –2LL | AIC | DIC | Time |
|---|---|---|---|---|---|---|
| DINA | 0.553 | 91 | 5451.89 | 5542.89 | 6336.40 | 1534.97 |
| rRUM | 0.701 | 107 | 4843.06 | 4950.06 | 6302.98 | 3369.94 |

*Note*, NP = number of estimated parameters; –2LL = –2 log likelihood; AIC = Congdon's version of Akaike's information criterion; DIC = deviance information criterion; Time = overall computing time for two Markov chains (in seconds).

**Table 13**. Estimates of Item Parameters in the Empirical Example.

| Item | DINA | | rRUM | | | | | |
|---|---|---|---|---|---|---|---|---|
| | $g$ | $s$ | $\pi^*$ | $r_1^*$ | $r_2^*$ | $r_3^*$ | $r_4^*$ | $r_5^*$ |
| $\frac{3}{4}-\frac{3}{8}$ | 0.011 (0.012) | 0.277 (0.024) | 0.894 (0.020) | 0.019 (0.016) | | | | |
| $3\frac{1}{2}-2\frac{3}{2}$ | 0.214 (0.025) | 0.121 (0.021) | 0.890 (0.020) | 0.539 (0.107) | 0.861 (0.106) | 0.842 (0.113) | 0.254 (0.050) | |
| $\frac{6}{7}-\frac{4}{7}$ | 0.146 (0.053) | 0.039 (0.010) | 0.969 (0.010) | 0.498 (0.039) | | | | |
| $3-2\frac{1}{5}$ | 0.127 (0.019) | 0.137 (0.029) | 0.894 (0.030) | 0.039 (0.040) | 0.806 (0.153) | 0.608 (0.212) | 0.769 (0.119) | 0.150 (0.079) |
| $3\frac{7}{8}-2$ | 0.217 (0.063) | 0.249 (0.023) | 0.746 (0.023) | | | 0.419 (0.063) | | |
| $4\frac{4}{12}-2\frac{7}{12}$ | 0.036 (0.012) | 0.229 (0.027) | 0.801 (0.027) | 0.110 (0.083) | 0.421 (0.211) | 0.645 (0.219) | 0.114 (0.043) | |
| $4\frac{1}{3}-2\frac{4}{3}$ | 0.076 (0.016) | 0.080 (0.018) | 0.939 (0.016) | 0.614 (0.151) | 0.314 (0.177) | 0.527 (0.226) | 0.116 (0.036) | |
| $\frac{11}{8}-\frac{1}{8}$ | 0.174 (0.045) | 0.051 (0.014) | 0.948 (0.014) | 0.940 (0.050) | 0.060 (0.046) | | | |
| $3\frac{4}{5}-3\frac{2}{5}$ | 0.108 (0.037) | 0.062 (0.014) | 0.957 (0.012) | 0.786 (0.058) | | 0.093 (0.048) | | |
| $2-\frac{1}{3}$ | 0.170 (0.023) | 0.075 (0.021) | 0.930 (0.023) | 0.132 (0.068) | | 0.338 (0.152) | 0.738 (0.106) | 0.362 (0.088) |
| $4\frac{5}{7}-1\frac{4}{7}$ | 0.123 (0.035) | 0.102 (0.017) | 0.915 (0.017) | 0.843 (0.062) | | 0.073 (0.039) | | |
| $7\frac{3}{5}-\frac{4}{5}$ | 0.034 (0.013) | 0.137 (0.022) | 0.881 (0.022) | 0.661 (0.170) | | 0.026 (0.028) | 0.088 (0.032) | |
| $4\frac{1}{10}-2\frac{8}{10}$ | 0.137 (0.021) | 0.161 (0.024) | 0.856 (0.023) | 0.374 (0.123) | 0.320 (0.230) | 0.160 (0.157) | 0.447 (0.070) | |
| $4-1\frac{4}{3}$ | 0.025 (0.010) | 0.203 (0.033) | 0.820 (0.036) | 0.090 (0.101) | 0.313 (0.260) | 0.276 (0.244) | 0.228 (0.079) | 0.120 (0.075) |
| $4\frac{1}{3}-1\frac{5}{3}$ | 0.015 (0.007) | 0.185 (0.025) | 0.838 (0.024) | 0.693 (0.180) | 0.112 (0.131) | 0.256 (0.198) | 0.015 (0.014) | |

*Note*, posterior standard deviations (i.e., standard errors) in parentheses.

**Table 14**. Estimates of Mixing Proportions in the Empirical Example.

| Pai[i] | Attribute Patterns | DINA | rRUM |
|---|---|---|---|
| pai[1] | 0 0 0 0 0 | 0.017 (0.012) | 0.032 (0.025) |
| pai[2] | 1 0 0 0 0 | 0.009 (0.007) | 0.005 (0.005) |
| pai[3] | 0 1 0 0 0 | 0.017 (0.015) | 0.006 (0.005) |
| ... | ... | ... | ... |
| pai[30] | 1 0 1 1 1 | 0.005 (0.004) | 0.003 (0.003) |
| pai[31] | 0 1 1 1 1 | 0.009 (0.009) | 0.010 (0.008) |
| pai[32] | 1 1 1 1 1 | 0.350 (0.022) | 0.328 (0.024) |

*Note*, posterior standard deviations (i.e., standard errors) in parentheses; the middle 26 patterns are omitted.

## Summary

This paper presents a systematic introduction to using **JAGS** for Bayesian CDM estimation. Several **JAGS** code are presented to fit some common CDMs. The unstructured latent structural model and the higher-order latent structural model are both introduced. It further demonstrates how to extend these models to the polytomous attributes, the testlet effect, and the longitudinal data. Finally, an empirical example is presented to illustrate how to utilize the **R2jags** package to run the **JAGS** code.

As a tutorial, this paper has its limitations. First, there is no way to exhaust all CDMs. Thus, the readers are encouraged to consult other sources for further reading. Second, some emerging research topics are not included, such as the Q-matrix estimation (e.g., Chen, Culpepper, Chen, & Douglas, 2018; Chung & Johnson, 2018), joint CDMs for response accuracy and response times (e.g., Zhan et al., 2018), and CDMs for polytomous scoring items (e.g., Ma & de la Torre, 2016; von Davier, 2008). Third, only the **R2jags** package was used to call **JAGS**, some other **R** packages like the **rjags** (Plummer, Stukalov, & Denwood, 2016) and **jagsUI** (Kellner, 2017) also can be used. Fourth, the computing time could be very long, especially for large-scale tests with a large sample size. Thus, it is desirable to develop more effective Bayesian estimation programs to increase the efficiency in model parameter estimation for the new models, such as the **dina** (Culpepper, 2015b) package in **R**. Additionally, a new Bayesian software, **Stan** (Carpenter et al., 2016), has been developed. **Stan** uses the no-U-turn sampler (Hoffman & Gelman, 2014), an extension to the Hamiltonian Monte Carlo (Neal, 2011) algorithm. Hamiltonian Monte Carlo is considerably faster than the Gibbs sampler which is used in **JAGS**. Further exploration would be valuable to use **Stan** to fit Bayesian CDMs (e.g., Lee, 2016).

All in all, given the increasing number of applications of the Bayesian MCMC algorithm in fitting many CDMs, **JAGS** can become a popular tool in the field. It is hoped that researchers can adapt the codes presented in this paper for their own testing situations and applied assessment challenges for cognitive diagnosis.